\documentclass[conference]{IEEEtran}
\IEEEoverridecommandlockouts 

\usepackage{cite}

\usepackage{subcaption}

\ifCLASSINFOpdf
   \usepackage[pdftex]{graphicx}
   \graphicspath{{./}}
   \DeclareGraphicsExtensions{.pdf,.jpeg,.png}
\else
   \usepackage[dvips]{graphicx}
   \graphicspath{{../eps/}}
   \DeclareGraphicsExtensions{.eps}
\fi
\usepackage[cmex10]{amsmath}
\usepackage{amsmath}
\usepackage{color}
\usepackage{graphicx}
\usepackage{caption}
\usepackage{subcaption}
\usepackage{wrapfig}
\usepackage{epstopdf}

\usepackage[hidelinks]{hyperref}
\usepackage{multirow}
\usepackage{booktabs,colortbl}
\usepackage{epstopdf}
  \renewcommand\footnoterule{\vspace*{-3pt}%
     \hrule width 2in height 0.4pt
     \vspace*{2.6pt}}

\begin{document}
%
\title{Model Predictive Control of Voltage Source Converter in a HVDC System}

\author{%

\IEEEauthorblockN{Mohammad Amin and Marta Molinas}
\IEEEauthorblockA{Dept. of Engineering Cybernetics\\Norwegian University of Science and Technology\\Trondheim-7491, Norway\\mohammad.amin@ntnu.no}

%

\thanks{Identify applicable sponsor/s here. \emph{(sponsors)}}

}


\maketitle

\begin{abstract}
Model Predictive Control (MPC) method is a class of advanced control techniques most widely applied in industry. The major advantages of the MPC are its straightforward procedure which can be applied for both linear and nonlinear system. This paper proposes the use of MPC for voltage source converter (VSC) in a high voltage direct current (HVDC) system. A MPC controller is modeled based on the state-space model of a single VSC-HVDC station including the dynamics of the main ac grid. A full scale nonlinear switching model of point-to-point connected VSC-based HVDC system is developed in matlab/simulink association with SimPower system to demonstrate the application of the proposed controller. \\
\end{abstract}

\begin{IEEEkeywords}
VSC control, HVDC, Model predictive control.
\end{IEEEkeywords}

\section{Introduction}
As Voltage source converter (VSC) based high voltage direct current (HVDC) transmission system offers many advantages such as independent and fast control on active and reactive power, improving power quality, and feeding of remote isolated loads \cite{YifanZhu2012}-\cite{NguyenMau2011}, it’s required the advanced control techniques to achieve the precise control of active and reactive power flow to maintain the system stability, ensure robust operation, and high levels of efficiency. A wide range of studies on modeling and control of VSC-based HVDC systems have been published in last few years \cite{JBeertenTPS2014}-\cite{SCole2010} which also include the small-signal stability analysis \cite{MehdiPSCC2014}-\cite{AminRPG2014}. Several control technique have been studied and applied in stationery \cite{JCVasquez2011} and synchronous reference frame \cite{WangYan2009} to control the three phase power converter. The VSC operating with vector control strategy can perform the independent control by means of controlling real d-axis and imaginary q-axis current components. Outer loop proportional integral (PI) controller is utilized to achieve dc voltage or active power control by controlling the d-axis current components, and the ac voltage or reactive power control is achieved by controlling q-axis current components. The controller can be easily tuned from the transfer function of the converter dynamics by applying symmetrical optimum and modulus optimum criteria to achieve maximum flatness from the bode plot of the transfer function \cite{CBajracharya2008}, however stability problem arises when it needs to tune the multi-terminal HVDC (MT-HVDC) system. In this case model predictive control shows great potential to achieve desire performance in case of MT-HVDC system. \\
MPC method has been widely applied in power converters \cite{SKouro2009} because of having high dynamic performance, simple implementation, high flexibility  by implementing several objective function \cite{PCortes2008}-\cite{KAhmed2011}, and being capable to directly manage the ac signal without frequency transformation\cite{MAPerez2008}-\cite{PCortes2010}. \\
\begin{figure*}[t] 
    \center
    \includegraphics[width=1\textwidth]{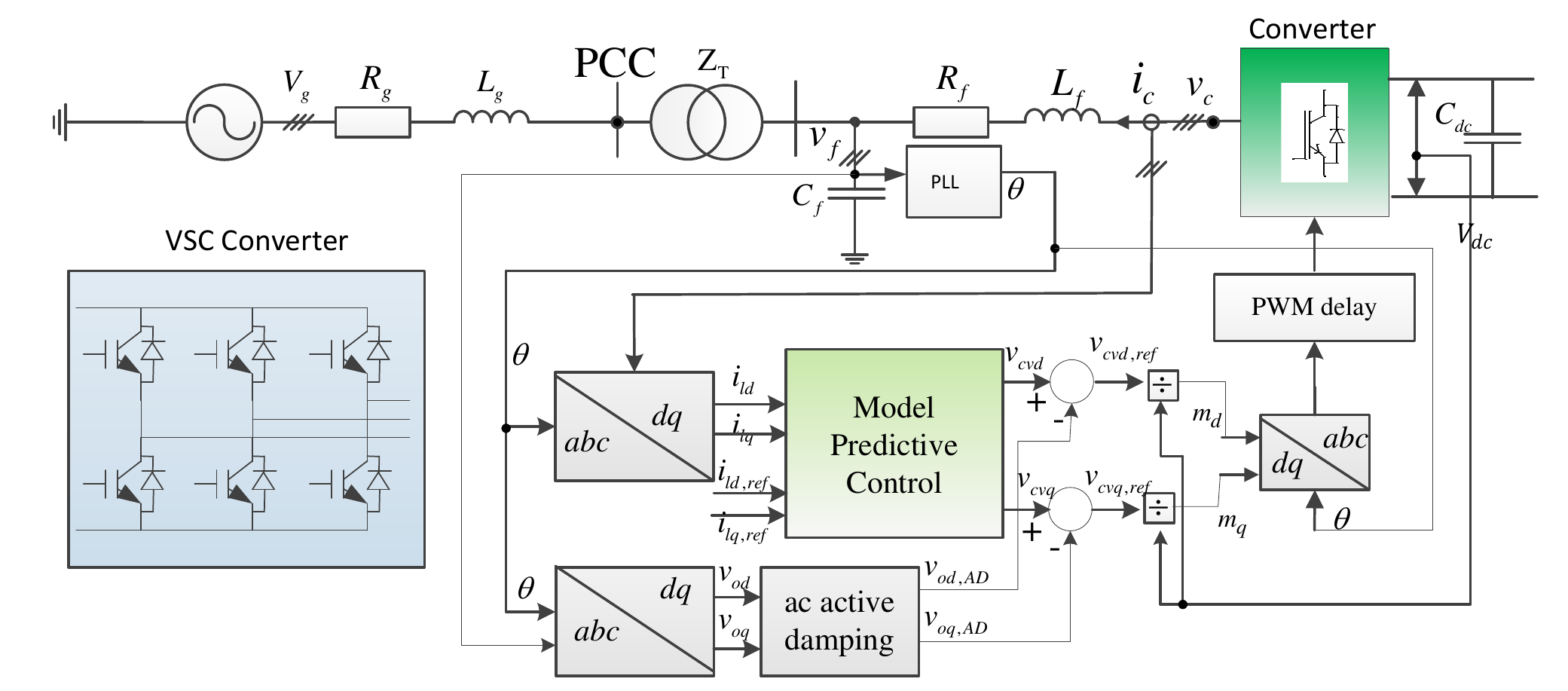}
\caption{Overview of single VSC-Converter station including the grid dynamics and MPC controller}
\label{fig:1}
\end{figure*} 
In this paper, MPC method is proposed to control the VSC in a HVDC system. The MPC is based on the prediction of the system response to a change in control variables in order to attain a minimum error. It performs following three steps in each control interval; 1) predict the future behavior of the system, 2) perform the optimization algorithm to calculate the future inputs for having desired output, and 3) process a certain number of calculated inputs \cite{EFCamacho1995}. The performance of the proposed method is investigated for linearized model of a single VSC-based HVDC system and finally a switching model of point-to-point VSC HVDC system is developed to demonstrate the application of this proposed method. 

\section{Analytical Modeling of VSC-HVDC system}

The electrical circuit of an inverter for analytical modeling is shown in fig. \ref{fig:1} where L$_c$ and R$_c$ are the total series inductance and resistance between the inverter and point of common coupling (PCC) and $C_{dc}$ is the dc link capacitor. $C_f$ is the filter capacitance connected at PCC, and $R_g$ and $L_g$ are grid resistance and inductance including series resistance and inductance of the transformer. The dynamic equations of the inverter in per unit (pu) can be given by (\ref{eqn:1}) and (\ref{eqn:2}), filter by (\ref{eqn:3cf}) and grid by (\ref{eqn:4})  where $\omega$$_ {b}$ is base angular grid frequency; $\omega$$_ {g}$ is grid frequency in pu; voltage and current of these equations is indicated in fig. \ref{fig:1}   \cite{VBlasko1997},\cite{NKroutikova2007}.  
\begin{equation}
\label{eqn:1}
\frac{d{{i}_{L}}}{dt}=\frac{{{\omega }_{b}}}{{{L}_{c}}}{{v}_{cv}}-\frac{{{\omega }_{b}}}{{{L}_{c}}}{{v}_{o}}-{{\omega }_{b}}(\frac{{{R}_{c}}}{{{L}_{c}}}+j{{\omega }_{g}}){{i}_{L}}
\end{equation}
\begin{equation}
\label{eqn:2}
\frac{d\,{{v}_{dc}}}{dt}=\frac{{{\omega }_{b}}}{{{C}_{dc}}}{{i}_{dc,line}}-\frac{{{\omega }_{b}}}{{{C}_{dc}}}{{i}_{dc}}
\end{equation}
\begin{equation}
\label{eqn:3cf}
\frac{d{{v}_{o}}}{dt}=\frac{{{\omega }_{b}}}{{{C}_{f}}}{{i}_{L}}-\frac{{{\omega }_{b}}}{{{C}_{f}}}{{i}_{o}}-j{{\omega }_{b}}{{\omega }_{g}}{{v}_{o}}
\end{equation}
\begin{equation}
\label{eqn:4}
\frac{d{{i}_{o}}}{dt}=\frac{{{\omega }_{b}}}{{{L}_{g}}}{{v}_{o}}-\frac{{{\omega }_{b}}}{{{L}_{g}}}{{v}_{g}}-{{\omega }_{b}}(\frac{{{R}_{g}}}{{{L}_{g}}}+j{{\omega }_{g}}){{i}_{o}}
\end{equation}
\begin{figure}[!t] 
    \center
    \includegraphics[width=.5\textwidth]{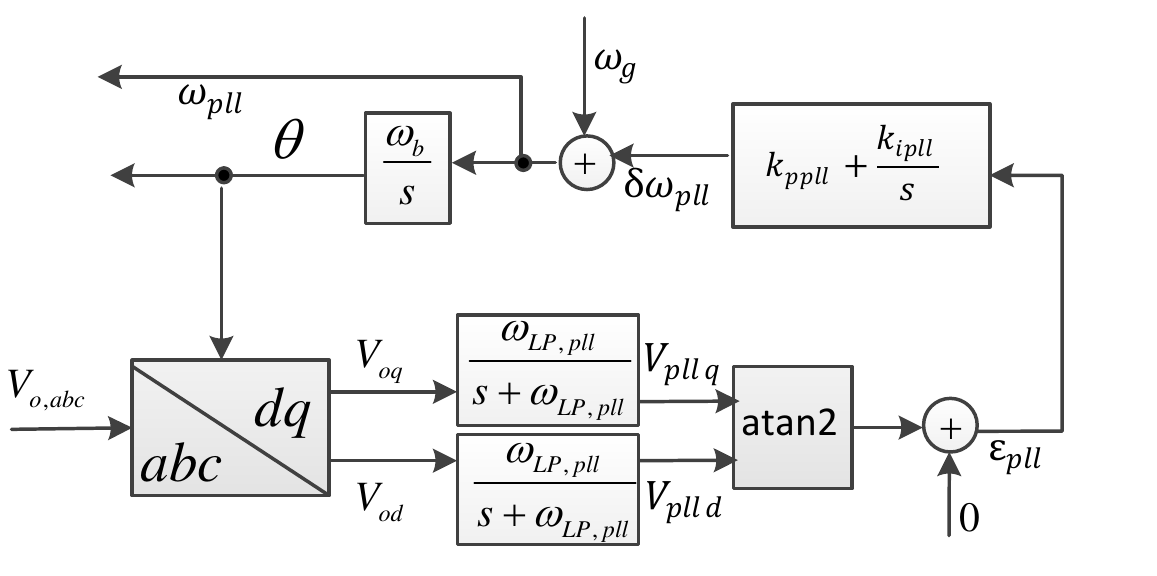}
\caption{Overview of implemented phase locked loop}
\label{fig:pll}
\end{figure}
\subsection{Phase Locked Loop}
The phase locked loop (PLL) is used to track the actual grid frequency \cite{VKaura1997}. An inverse tangent function is used on first order low-pass filtered output of the q- and d-axis voltage to estimate the actual phase angle error as shown in Fig. \ref{fig:pll}. This phase angle error is the input to the PI controller for tracking the frequency of the measured voltage. The states of the low-pass filter of the PLL are given by (\ref{eqn:pll1}) where $\omega$$_{LP,PLL}$  is the cut-off frequency of the low-pass filter.
\begin{subequations}
\label{eqn:pll1}
\begin{align}
\frac{d{{v}_{PLL,d}}}{dt}=-{{\omega }_{LP,PLL}}{{v}_{PLL,d}}+{{\omega }_{LP,PLL}}{{v}_{od}} \\ 
\label{eqn:19b}
\frac{d{{v}_{PLL,q}}}{dt}=-{{\omega }_{LP,PLL}}{{v}_{PLL,q}}+{{\omega }_{LP,PLL}}{{v}_{oq}}
\end{align}
\end{subequations}
The speed deviation, $\delta$$\omega$$_{PLL}$ of the PLL with respect to the grid frequency can be defined by (\ref{eqn:pll2}) where K$_{pPLL}$ and K$_{iPLL}$ are the proportional and integral gain of the PI controller; $\varepsilon$$_{PLL}$ is the variable introduced to represent the integrator state and can be defined by (\ref{eqn:pll3}). 
\begin{equation}
\label{eqn:pll2}
\delta {{\omega }_{PLL}}={{K}_{pPLL}}\arctan \left( \frac{{{v}_{PLL,q}}}{{{v}_{PLL,d}}} \right)+{{K}_{iPLL}}{{\varepsilon }_{PLL}}
\end{equation}
\begin{equation}
\label{eqn:pll3}
\frac{d\,{{\varepsilon }_{PLL}}}{dt}=\arctan \left( \frac{{{v}_{PLL,q}}}{{{v}_{PLL,d}}} \right)
\end{equation}
The corresponding phase angle, $\delta$$\theta$$_{PLL}$ difference between the grid voltage and orientation of the PLL is represented by (\ref{eqn:pll4}) and grid voltage represented by its amplitude $\hat{V}$ can be transformed into PLL reference frame as given by (\ref{eqn:pll5}). Frequency of the PLL can be given by (\ref{eqn:pll6}). 
\begin{align}
\label{eqn:pll4}
\frac{d\delta {{\theta }_{PLL}}}{dt}&=\delta {{\omega }_{PLL}}{{\omega }_{b}} ={{\omega }_{b}}{{K}_{pPLL}}\arctan \left( \frac{{{v}_{PLL,q}}}{{{v}_{PLL,d}}} \right)\nonumber \\ &+{{\omega }_{b}}{{K}_{iPLL}}{{\varepsilon }_{PLL}}
\end{align}
\begin{equation}
\label{eqn:pll5}
{{v}_{g}}={{{\hat{v}}}_{g}}{{e}^{-j\delta {{\theta }_{PLL}}}}
\end{equation}
\begin{equation}
\label{eqn:pll6}
{{\omega }_{PLL}}=\delta {{\omega }_{PLL}}+{{\omega }_{g}}
\end{equation}

\subsection{State-space realization}
The modeling, analysis and control of the system will be presented in a synchronous reference frame (SRF). The transformation of the three phase quantity from stationary reference frame to SRF is based on the amplitude-invariant Park transformation, with the d-axis aligned with the voltage vector $v_o$ and q-axis leading the d-axis by $90^0$. An ideal lossless average model is assumed for the converter. Therefore power balance constraint between dc and ac side can be given by 
\begin{equation}
\label{eqn:3}
{{i}_{dc}}{{v}_{dc}}={{i}_{Ld}}{{v}_{cvd}}+{{i}_{Lq}}{{v}_{cvq}}.
\end{equation}
In SRF the average model of the inverter, voltage over the filter capacitance and current of the grid inductance can be presented by (\ref{eqn:avrgmc1})-(\ref{eqn:avrgmc2}), (\ref{eqn:cfsrf}) and (\ref{eqn:lgsrf}), respectively.
\begin{subequations}
\label{eqn:avrgmc1}
\begin{align}
\label{eqn:6a}
\frac{d{{i}_{Ld}}}{dt}=\frac{{{\omega }_{b}}}{{{L}_{c}}}{{v}_{cvd}}-\frac{{{\omega }_{b}}}{{{L}_{c}}}{{v}_{od}}-\frac{{{\omega }_{b}}{{R}_{c}}}{{{L}_{c}}}{{i}_{Ld}}+{{\omega }_{b}}{{\omega }_{g}}{{i}_{Lq}} \\ 
\label{eqn:6b}
 \frac{d{{i}_{Lq}}}{dt}=\frac{{{\omega }_{b}}}{{{L}_{c}}}{{v}_{cvq}}-\frac{{{\omega }_{b}}}{{{L}_{c}}}{{v}_{oq}}-{{\omega }_{b}}{{\omega }_{g}}{{i}_{Ld}}-\frac{{{\omega }_{b}}{{R}_{c}}}{{{L}_{c}}}{{i}_{Lq}}
\end{align}
\end{subequations}
\begin{equation}
\label{eqn:avrgmc2}
\frac{d{{v}_{dc}}}{dt}=\frac{{{\omega }_{b}}}{{{C}_{dc}}}{{i}_{dc,line}}-\frac{{{\omega }_{b}}({{i}_{Ld}}{{v}_{cvd}}+{{i}_{Lq}}{{v}_{cvq}})}{{{C}_{dc}}{{v}_{dc}}}
\end{equation}
\begin{subequations}
\label{eqn:cfsrf}
\begin{align}
\label{eqn:19a}
\frac{d{{v}_{od}}}{dt}=\frac{{{\omega }_{b}}}{{{C}_{f}}}{{i}_{Ld}}-\frac{{{\omega }_{b}}}{{{C}_{f}}}{{i}_{od}}+{{\omega }_{b}}{{\omega }_{g}}{{v}_{oq}}  \\ 
\label{eqn:19b}
\frac{d{{v}_{oq}}}{dt}=\frac{{{\omega }_{b}}}{{{C}_{f}}}{{i}_{Lq}}-\frac{{{\omega }_{b}}}{{{C}_{f}}}{{i}_{oq}}-{{\omega }_{b}}{{\omega }_{g}}{{v}_{od}}
\end{align}
\end{subequations}
\begin{subequations}
\label{eqn:lgsrf}
\begin{align}
\label{eqn:19a}
\frac{d{{i}_{od}}}{dt}=\frac{{{\omega }_{b}}}{{{L}_{g}}}{{v}_{od}}-\frac{{{\omega }_{b}}}{{{L}_{g}}}{{{\hat{v}}}_{g}}\cos (\delta {{\theta }_{PLL}})-\frac{{{\omega }_{b}}{{R}_{g}}}{{{L}_{g}}}{{i}_{od}}+{{\omega }_{b}}{{\omega }_{g}}{{i}_{oq}}  \\ 
\label{eqn:19b}
\frac{d{{i}_{oq}}}{dt}=\frac{{{\omega }_{b}}}{{{L}_{g}}}{{v}_{oq}}+\frac{{{\omega }_{b}}}{{{L}_{g}}}{{{\hat{v}}}_{g}}sin(\delta {{\theta }_{PLL}})-{{\omega }_{b}}{{\omega }_{g}}{{i}_{od}}-\frac{{{\omega }_{b}}{{R}_{g}}}{{{L}_{g}}}{{i}_{oq}}
\end{align}
\end{subequations}
The average model of converter presented by (\ref{eqn:avrgmc1}) and (\ref{eqn:lgsrf}) is nonlinear and the nonlinearity prevents direct application of classical linear analysis techniques. Therefore, the model is linearized by taking first-order partial derivatives in respect to all variables in steady-state operating point. The states, inputs and output vector are given by (\ref{eqn:6}) and resulting state-space matrices are presented by (\ref{eqn:7}).
\begin{align}
\label{eqn:6}
x=&
\left[
\begin{matrix}
  i_{ld} & i_{lq}&  v_{dc}& i_{od}&i_{oq}&v_{od}&v_{oq} 
 \end{matrix}
 \right]^T  \nonumber \\
u=&
\left[
\begin{matrix}
 v_{cvd} & v_{cvq}&  v_{g}&  i_{dc} 
 \end{matrix}
 \right]^T \nonumber \\
y=&
\left[
\begin{matrix}
  i_{ld} & i_{lq} 
 \end{matrix}
 \right]^T 
\end{align}
A=\[\left( \begin{matrix}
   -\frac{{{\omega }_{b}}\,{{R}_{f}}}{{{L}_{f}}} & {{\omega }_{b}} & 0 & 0 & 0 & -\frac{{{\omega }_{b}}\,}{{{L}_{f}}} & 0  \\
   -{{\omega }_{b}} & -\frac{{{\omega }_{b}}\,{{R}_{f}}}{{{L}_{f}}} & 0 & 0 & 0 & 0 & -\frac{{{\omega }_{b}}\,}{{{L}_{f}}}  \\
   -\frac{{{D}_{d}}\,{{\omega }_{b}}}{{{C}_{dc}}} & -\frac{{{D}_{q}}{{\omega }_{b}}}{{{C}_{dc}}} & C_{3,3} & 0 & 0 & 0 & 0  \\
   0 & 0 & 0 & -\frac{{{\omega }_{b}}{{R}_{g}}}{{{L}_{g}}} & {{\omega }_{b}} & \frac{{{\omega }_{b}}}{{{L}_{g}}} & 0  \\
   0 & 0 & 0 & -{{\omega }_{b}} & -\frac{{{\omega }_{b}}{{R}_{g}}}{{{L}_{g}}} & 0 & \frac{{{\omega }_{b}}}{{{L}_{g}}}  \\
   \frac{{{\omega }_{b}}}{{{C}_{f}}} & 0 & 0 & -\frac{{{\omega }_{b}}}{{{C}_{f}}} & 0 & 0 & {{\omega }_{b}}  \\
   0 & \frac{{{\omega }_{b}}}{{{C}_{f}}} & 0 & 0 & -\frac{{{\omega }_{b}}}{{{C}_{f}}} & -{{\omega }_{b}} & 0  \\
\end{matrix} \right) \]
Element of A matrix, C$_{3,3}$ is as follows.
\[C_{3,3}=\frac{{{\omega }_{b}}\left( {{D}_{d}}{{I}_{ld0}}+{{D}_{q}}{{I}_{lq0}} \right)}{{{C}_{dc}}{{V}_{dc0}}}\]

\begin{align}
\label{eqn:7}
B&
=
\left( \begin{matrix}
   \frac{{{\omega }_{b}}}{{{L}_{f}}} & 0 & 0 & 0  \\
   0 & \frac{{{\omega }_{b}}}{{{L}_{f}}} & 0 & 0  \\
   -\frac{{{D}_{d}}{{\omega }_{b}}}{{{C}_{dc}}{{V}_{dc0}}} & -\frac{{{D}_{q}}{{\omega }_{b}}}{{{C}_{dc}}{{V}_{dc0}}} & 0 & \frac{{{\omega }_{b}}}{{{C}_{dc}}}  \\
   0 & 0 & -\frac{{{\omega }_{b}}\cos \left( \delta {{\theta }_{PLL}} \right)}{\lg } & 0  \\
   0 & 0 & \frac{{{\omega }_{b}}\sin \left( \delta {{\theta }_{PLL}} \right)}{\lg } & 0  \\
   0 & 0 & 0 & 0  \\
   0 & 0 & 0 & 0  \\
\end{matrix} \right) \nonumber \\
C&
=
\left(
\begin{matrix}
1 & 0&0&0&0&0&0 \\
0& 1&0&0&0&0&0
 \end{matrix}
 \right) \nonumber \\
D&
=
\left(
\begin{matrix}
0 & 0 &0 &0 \\
0 & 0 &0 &0
 \end{matrix}
 \right) 
\end{align}
The grid voltage is assumed stable and a constant value and the dc current depends on output current $i_{Ld}$. Neglecting these two variables, the input vector and B matrix can be written by 
\begin{align}
u=&
\left[
\begin{matrix}
 v_{cvd} & v_{cvq} 
 \end{matrix}
 \right]^T \nonumber \\
B=&\left( \begin{matrix}
   \frac{{{\omega }_{b}}\,}{{{L}_{f}}} & 0  \\
   0 & \frac{{{\omega }_{b}}\,}{{{L}_{f}}}  \\
   -\frac{{{D}_{d}}\,{{\omega }_{b}}}{{{C}_{dc}}{{V}_{dc0}}} & -\frac{{{D}_{q}}{{\omega }_{b}}}{{{C}_{dc}}{{V}_{dc0}}}  \\
   0 & 0  \\
   0 & 0  \\
   0 & 0  \\
   0 & 0  \\
\end{matrix} \right).
\end{align}
The mapping between input and output variables can be solved in the Laplace domain by utilizing (\ref{eqn:18}) where matrix G(s) contains small signal transfer functions of the VSC converter at open loop and can be written by (\ref{eqn:19}).
\begin{equation}
\label{eqn:18}
Y(s)=[C(sI-A)^{-1}B+D] U(s)=G(s)U(s)
\end{equation}
\begin{subequations}
\label{eqn:19}
\begin{align}
i_{ld}(s)=&G_{d1}(s)V_{cvd}+G_{d2}(s)V_{cvq} \\
i_{lq}(s)=&G_{q1}(s)V_{cvd}+G_{q2}(s)V_{cvq}
\end{align}
\end{subequations}
\begin{figure}[t] 
    \center
    \includegraphics[width=.5\textwidth]{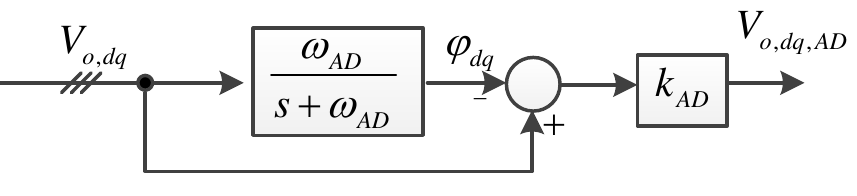}
\caption{Implemented active ac damping}
\label{fig:acdamp}
\end{figure} 
\begin{figure*}[t] 
    \center
    \includegraphics[width=1\textwidth]{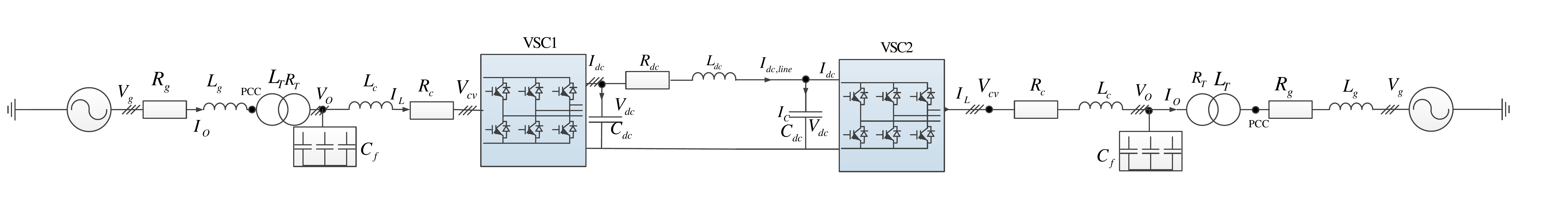}
\caption{Investigated point-to-point VSC-HVDc system}
\label{fig:p2p}
\end{figure*} 
\subsection{Active AC damping}
The active ac damping is designed to suppress the LC oscillations in the filter \cite{MoOlve2013}. There are several concepts developed for damping such oscillations; in this case the active damping, is based on injecting a voltage component of counter phase with detected oscillation in order to produce a cancellation effect. The oscillation is first isolated by high pass filtering and is then multiplied by a gain $K_{AD}$. The high pass filter function is implemented by subtracting from measure voltage  signals a low pass filtered version of same voltages as shown in Fig. \ref{fig:acdamp}. The damping voltage reference is given by (\ref{eqn:acdamp1}) where $\varphi$$_{dq}$ is the low pass filtered voltage signal of measure voltages signal. The corresponding internal states $\varphi$$_{d}$ and $\varphi$$_{q}$ of low pass filter can be given by (\ref{eqn:acdamp2}), where $\omega$$_{AD}$ is the cut-off frequency of the applied low-pass filter. 
\begin{equation}
\label{eqn:acdamp1}
{{v}_{o,dq,AD}}={{K}_{AD}}(-{{\varphi }_{dq}}+{{v}_{o,dq}})
\end{equation}
\begin{subequations}
\label{eqn:acdamp2}
\begin{align}
\label{eqn:19a}
\frac{d{{\varphi }_{d}}}{dt}=-{{\omega }_{AD}}{{\varphi }_{d}}+{{\omega }_{AD}}{{v}_{od}} \\ 
\label{eqn:19b}
\frac{d{{\varphi }_{q}}}{dt}=-{{\omega }_{AD}}{{\varphi }_{q}}+{{\omega }_{AD}}{{v}_{oq}}
\end{align}
\end{subequations}
The voltage reference to the converter can be written 
\begin{equation}
{{v}_{cvdq,ref}}(k+i)={{v}_{cvdq}}(k+i)-{{v}_{o,dq,AD}}(k)
\end{equation}
\begin{figure}[t] 
    \center
    \includegraphics[width=.54\textwidth]{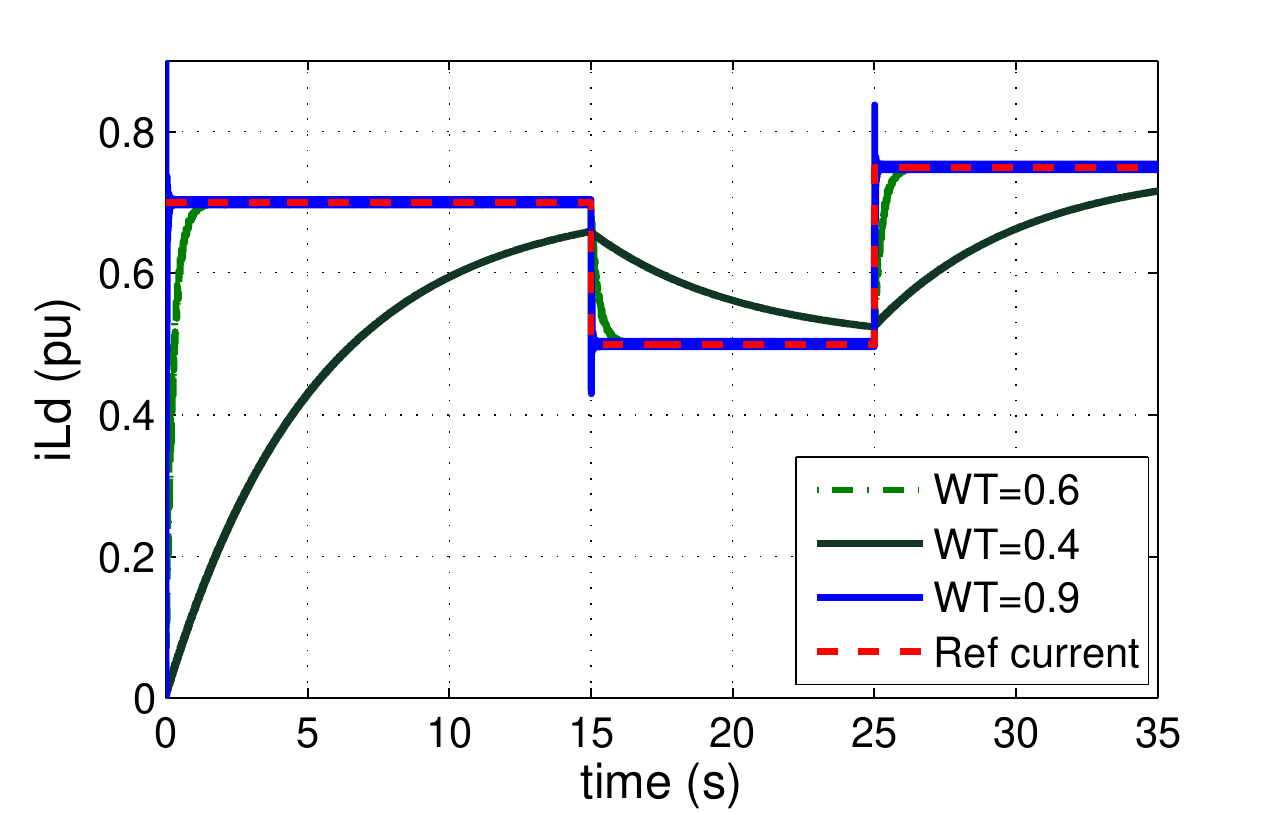}
\caption{Performance of the controller for different weight tuning.}
\label{fig:wte}
\end{figure} 
\section{Proposed Model Predictive Control}
The use of MPC to control the power of a VSC-based HVDC system is proposed in this study. The controller is designed based on the system presented in state-space form in previous section. The system is in continuous time. It is necessary to converter it into discrete time since MPC controller performs all the estimation and optimization calculation in discrete time. In discrete time it can be written by  
\begin{align}
\label{eqn:ssdis}
  & x(k+1)=A\,x(k)+B\,u(k) \nonumber \\ 
 & y(k+1)=C\,x(k)+D\,u(k).  
\end{align}
MPC solves an optimization problem at each control interval. The solutions determine the input variables to be used in the next control interval. A standard cost function is defined by 
\begin{align}
\label{eqn:cf}
J({{z}_{k}})=\sum\limits_{i=0}^{p}{[{{\{w[{{i}_{Ld,ref}}(k+i|k)-{{i}_{Ld}}(k+i|k)]\}}^{2}}} \nonumber \\ +{{\{w[{{i}_{Lq,ref}}(k+i|k)-{{i}_{Lq}}(k+i|k)]\}}^{2}}]+{{\rho }_{\varepsilon }}\varepsilon _{k}^{2}
\end{align}
where k is the current control interval; p is prediction horizon (number of intervals); w is the tuning weight; ${i}_{Ld,ref}(k+i|k)$ and ${i}_{Lq,ref}(k+i|k)$ are the predicted value of reference current at i-th prediction horizon step; ${i}_{Ld}(k+i|k)$ and ${i}_{Lq}(k+i|k)$ are the predicted value of output current at i-th prediction horizon step; $\varepsilon $$_{k}$ is the slack variable; $\rho$$_\varepsilon $$_k$ is the constraint violation penalty weights and $z_k$ is the decision taken by Quadric Program (QP) and can be given by 
 \begin{equation}
{{z}_{k}}=[u{{(k|k)}^{T}}\,\,\,u{{(k+1|k)}^{T}}\,\,...\,\,\,u{{(k+p-1|k)}^{T}}\,\,{{\varepsilon }_{k}}]
\end{equation}
and the input and output constraints are bound as follows:
\begin{align}
  & ~{{i}_{d,min}}(i)-{{\varepsilon }_{k}}V_{min}^{id}(i)\le {{i}_{Ld}}(k+i|k)\le {{i}_{d,max}}(i)+{{\varepsilon }_{k}}V_{min}^{id}(i)\nonumber\\ 
 & {{i}_{q,min}}(i)-{{\varepsilon }_{k}}V_{min}^{iq}(i)\le {{i}_{Lq}}(k+i|k)\le {{i}_{q,max}}(i)+{{\varepsilon }_{k}}V_{min}^{iq}(i)\nonumber\\ 
 & {{v}_{d,min}}(i)-{{\varepsilon }_{k}}V_{min}^{vd}(i)\le {{v}_{cvd}}(k+i|k)\le {{v}_{d,max}}(i)+{{\varepsilon }_{k}}V_{min}^{vd}(i)\nonumber\\ 
 & {{v}_{q,min}}(i)-{{\varepsilon }_{k}}V_{min}^{vq}(i)\le {{v}_{cvq}}(k+i|k)\le {{v}_{q,max}}(i)+{{\varepsilon }_{k}}V_{min}^{vq}(i)\nonumber \\  
\end{align}
where i=1:p and V$_{min}^{id}$, V$_{min}^{iq}$, V$_{min}^{vd}$ and V$_{min}^{vq}$ are the controller constants analogous to the cost functions weights which are used for the constraints softening.
\begin{table}[t]
\renewcommand{\arraystretch}{1.2}
%
\caption{Data used for the point-to-point HVDC system}
\label{tab:1}
\noindent
\centering
    \begin{minipage}{\linewidth} 
    \renewcommand\footnoterule{\vspace*{-5pt}} 
    \begin{center}
 \begin{tabular}{ | l | l | l |  l |}
    \hline
    Parameter & Value & Parameter & Value \\ \hline
     S$_{b}$ & 200 MVA &  L$_c$ & 0.15 pu\\ \hline
   V$_{ac}$ & 230 kV &  R$_c$ & 0.0015 pu \\ \hline  
$\omega$ & 2$\pi$50 rad/s &  C$_{f}$  & 0.094 pu \\ \hline
Transformer turn ratio &230/100 &V$_{dc}$ & 200 kV\\ \hline
L$_{T}$ & 0.15 pu & dc line length & 75 km\\ \hline
R$_{T}$ & 0.027 pu &L$_{dc}/km$ & 2.615 mH\\ \hline
L$_{g}$ & 0.0739 pu & R$_{dc}/km$ &0.011 $\Omega$  \\ \hline
R$_{g}$ & 0.0521 pu& C$_{dc}$ &4.2224 pu \\    \hline
        \end{tabular}
        \end{center}
    \end{minipage}
\end{table}
\begin{figure}[t] 
    \center
    \includegraphics[width=.5\textwidth]{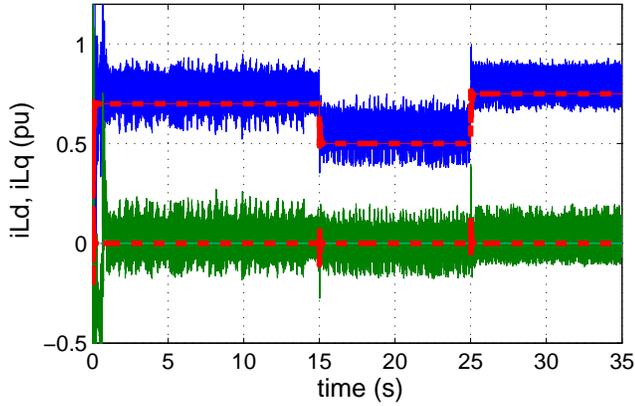}
\caption{d-axis current (upper curve) and q-axis current (lower curve). Initial current reference is 0.7 pu. At 15 s d-axis current reference steps down for 0.2 pu and at 25 s steps up for 0.25 pu. The q-axis current reference keeps 0 pu. (Green solid line for switching model and red dash line for state-space linearized model)}
\label{fig:ildq}
\end{figure}
\begin{figure}[t]
\begin{subfigure}[t]{.5\textwidth}
    \center
    \includegraphics[width=1\textwidth]{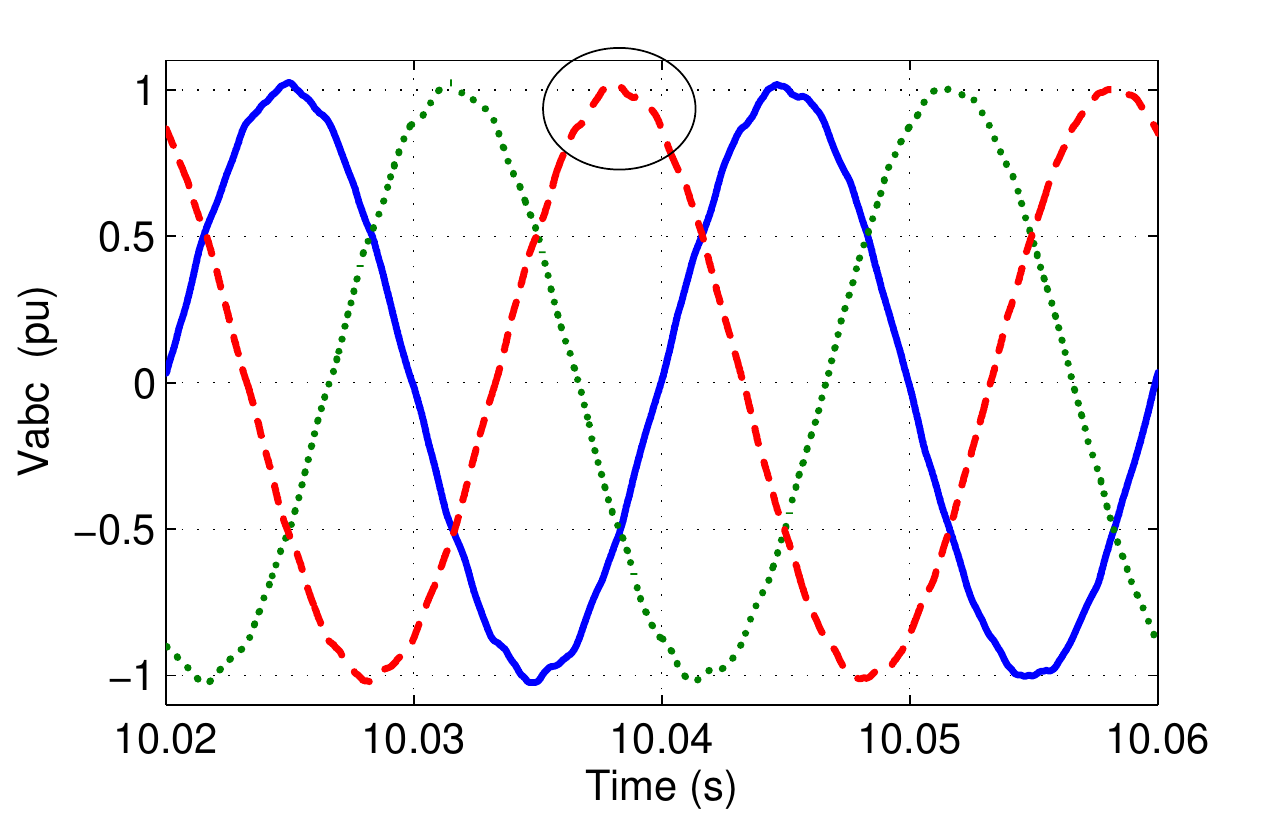}
\caption{}
\end{subfigure}
\begin{subfigure}[t]{.5\textwidth}
    \center
    \includegraphics[width=1\textwidth]{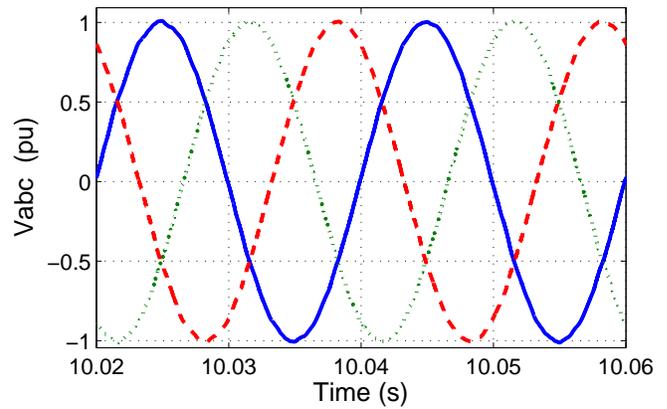}
\caption{}
\end{subfigure}
\caption{Three phase voltage Instantaneous voltage at interfacing point. (a) without ac active damping, (b) with ac active damping}
\label{fig:Vabc}
\end{figure} 
\section{Result and Analysis}

The investigated point-to-point VSC-based HVDC system is shown in fig. \ref{fig:p2p}. VSC1 is utilized to control the active and reactive power while VSC2 is used to control the dc link voltage and reactive power. The electrical parameters used for the system are given in table \ref{tab:1}. The converters are connected to the ac grid through a transformer of 230 kV/100 kV as same rating as converter. Model predictive control is used in VSC1 to control the active and reactive power by controlling the d-axis and q-axis current components of converter inductor, while for the VSC2, widely used decouple PI current control is used for inner current controller and PI controllers are used for outer loop dc voltage control and reactive power control. \\
\begin{figure}[!t] 
    \center
    \includegraphics[width=.53\textwidth]{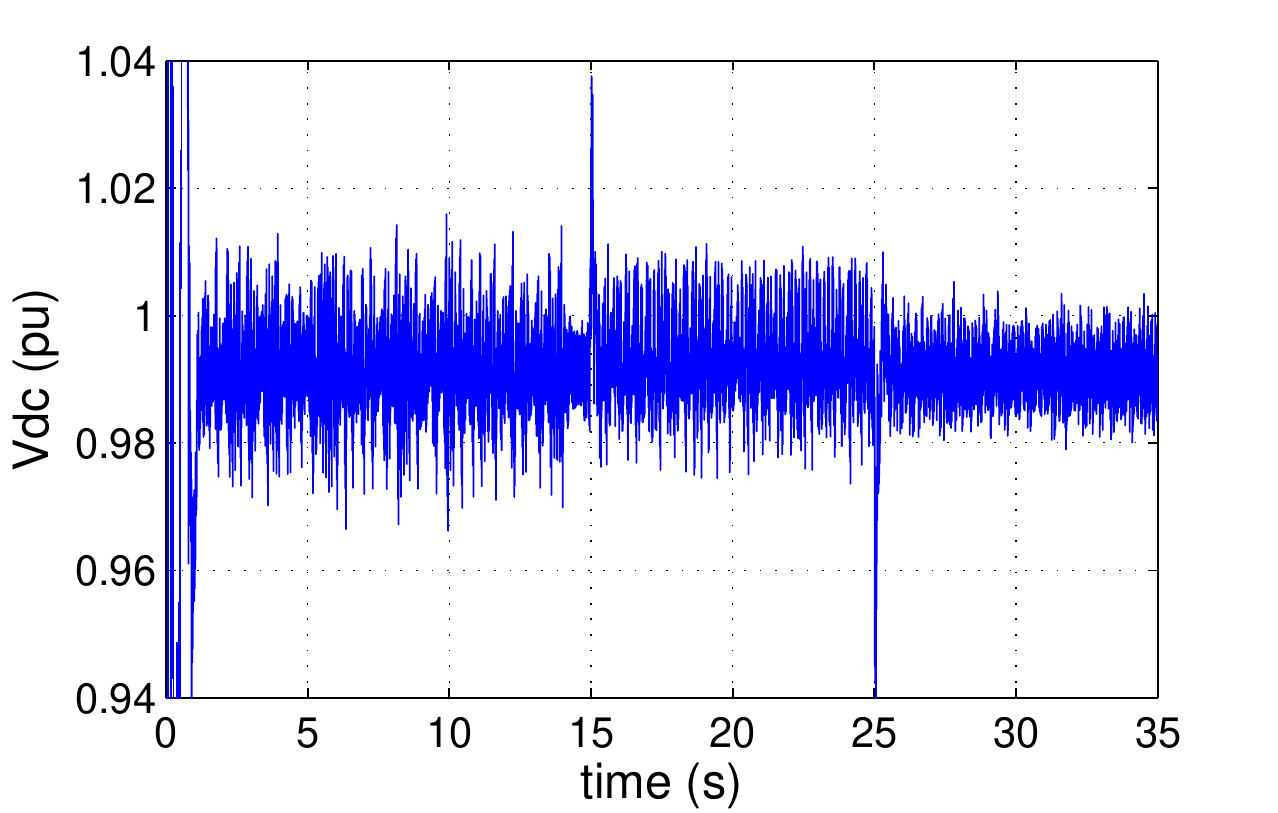}
\caption{dc voltage at VSC2 converter station from switching model}
\label{fig:dcvoltage}
\end{figure} 
\begin{figure}[!t] 
    \center
    \includegraphics[width=.53\textwidth]{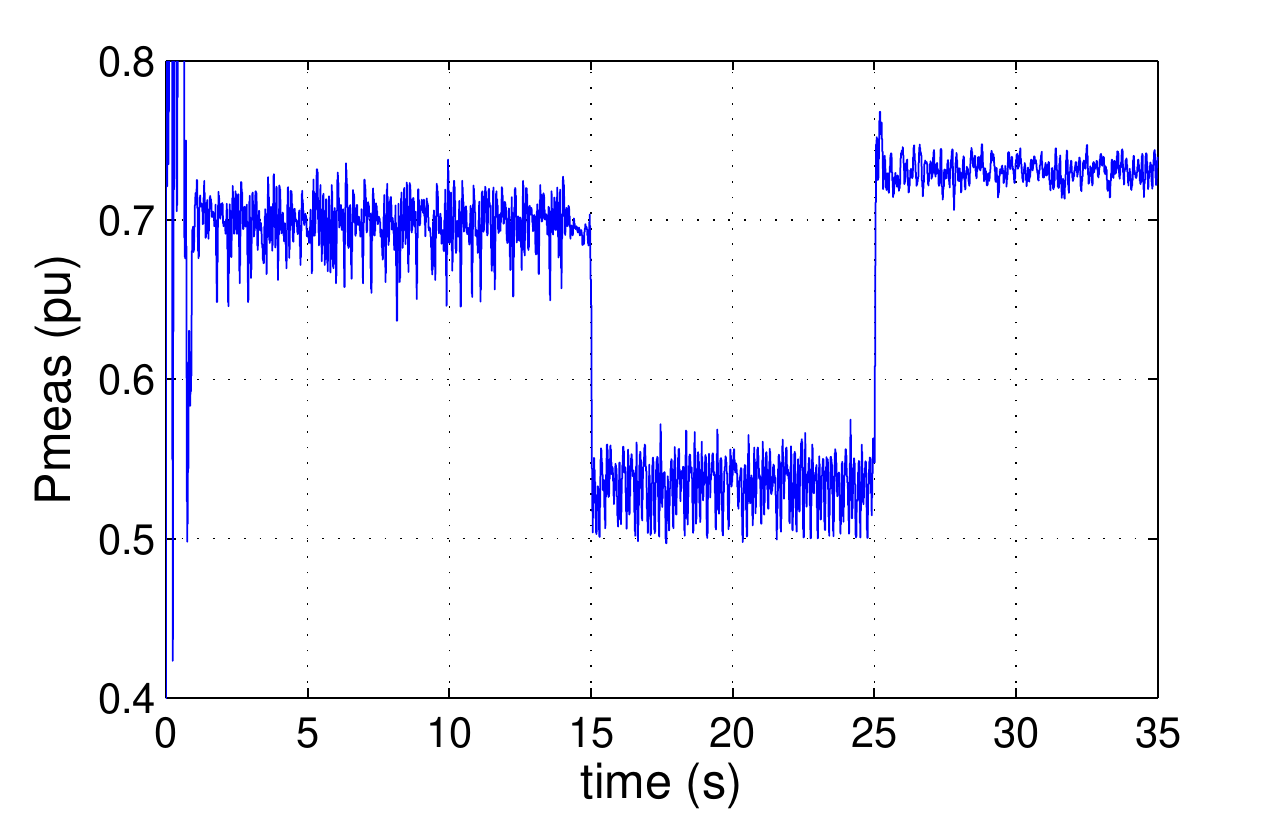}
\caption{active power of VSC2 converter from switching model}
\label{fig:pmeas}
\end{figure} 
\begin{figure}[!t] 
    \center
    \includegraphics[width=.53\textwidth]{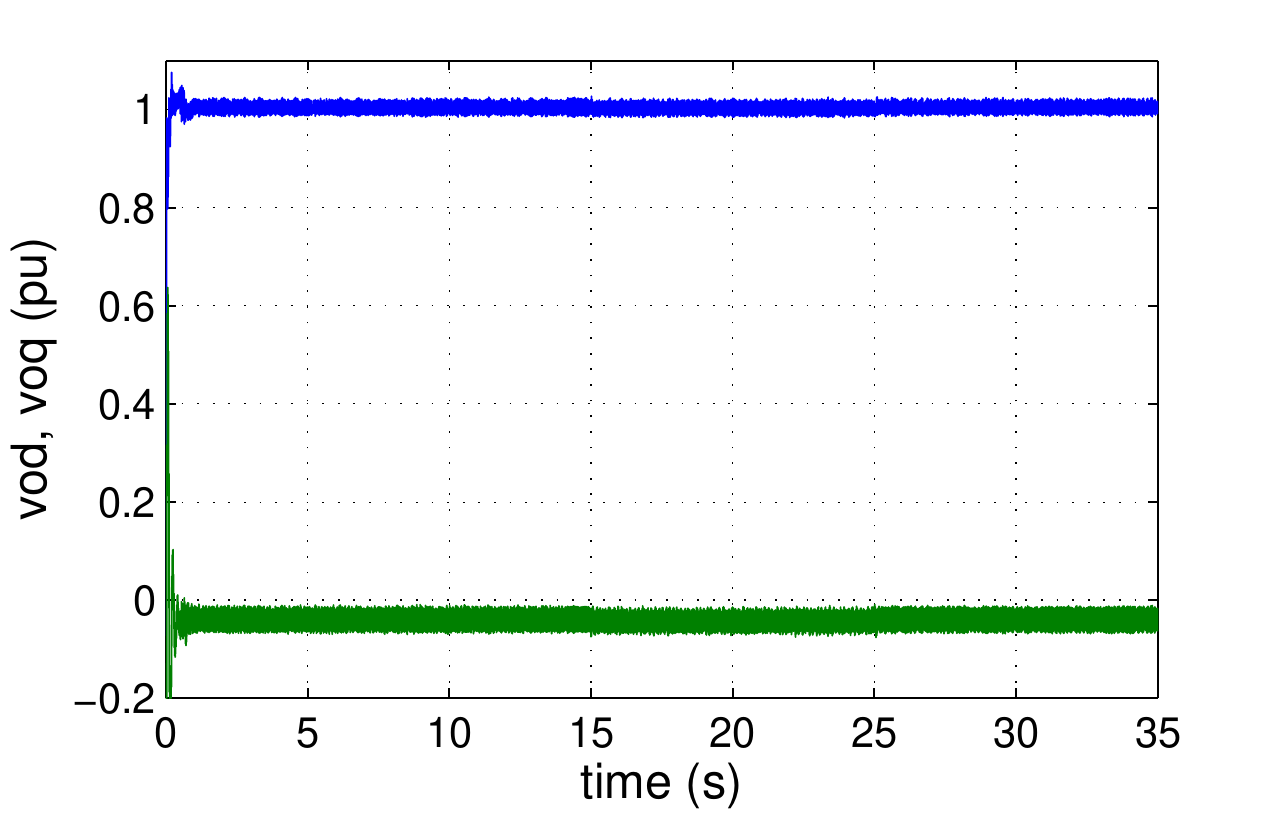}
\caption{d-axis and q-axis voltage components at filter capacitor connecting point}
\label{fig:vodvoq}
\end{figure}
The MPC controller for VSC1 is optimized for the plant model of VSC converter station presented in state-space form including the dynamics of the ac grid. The performance of the controller depends on appropriate selection of control parameters. The control parameters of the MPC include the sample time, T$_s$, Prediction horizon, p, Control horizon, m, Scale factor and tuning weights.\\  
 The performance of the modeled controller for different weight of tuning is shown in fig. \ref{fig:wte}. This result is obtained from time domain response of state-space model. At a lower tuning weights of 0.4 it takes longer time to follow the reference current as shown in fig. \ref{fig:wte} by cyan dash-dot curve. When it increase the tuning weight to .6, the performance of the system improves much better. If it increases to much higher value, for example here it is increase to 0.9, the time response becomes better, but it reduces the robustness of the system.\\ 
The modeled MPC is used to control active and reactive power by means of controlling the d-axis and q-axis current of the converter. The performance of the controller is validated by time domain response of state-space model and also from full scale nonlinear switching model of single VSC-HVDC converter in a point-to-point connection HVDC transmission system. \\
 The d-axis and q-axis current from state-space model and switching model is shown in fig. \ref{fig:ildq}. The initial d-axis current reference is set to 0.7 pu and the q-axis current reference is set to 0 pu. At 15 s, the d-axis current reference steps down for 0.2 pu and steps up for 0.25 pu at 25 s. Both state-space model and switching model can follow the reference. The high frequency components can be removed by using appropriate damping constant and cut-off frequency of active damping. Fig. \ref{fig:Vabc} shows the improvement of using ac active damping term by removing ripple components.  The dc voltage, measured active power and d-axis and q-axis voltage components of filter capacitor of VSC1 are shown in fig. \ref{fig:dcvoltage}, \ref{fig:pmeas} and \ref{fig:vodvoq}, respectively. The dc voltage is always stable for step changing of current and is in nominal operating range from 0.95 to 1.05 pu. The active power also follows the reference power. The system is working properly and it shows the great potential of using MPC controller to control VSC in a HVDC system.    
\section{Conclusion}
In this paper a MPC controller is proposed for controlling the VSC in a HVDC transmission system. A point-to-point connection HVDC system is developed to demonstrate the application of the proposed MPC controller. The MPC controller is applied to control the d-axis and q-axis current of a VSC in HVDC system and the performance is found to be satisfactory. The performance of the controller depends selection of optimum control parameters. The sample time balances between computational effort and controller performance. Small value of the sampling time increases the computational effort when the high value leads to information loss. The value of the sampling time should be selected such a way that it can capture the behavior of the system operating in the level with the fastest dynamics. 
A lager value of prediction horizon, p improves the stability of the closed loop system. If the desired response time of the system is T, then p is selected such that T is equal to p times T$_s$ (T=pT$_s$). Smaller value of control horizon, m allows to compute fewer variable in the QP at each control interval which increase computation speed and promotes an internally stable controller.  Another important parameter is weight tuning, w defined in the cost function in (\ref{eqn:cf}). The value of w can be varied from 0-1. Lower value of w makes the system robust however the system shows slower response time. It is the trade of between robustness and system response time.
\end{document}